\documentstyle[proceedings]{article} 

\input{psfig.sty}
\input epsf

\def\ss{\scriptscriptstyle}

\def\lra{\longrightarrow}
\newcommand{\be}{\begin{eqnarray}}
\newcommand{\ee}{\end{eqnarray}}
\newcommand{\nn}{\nonumber}

\def\o{\over}

\def\fav{{\bar f}(t)}

\def\ra{\rightarrow}

\title{EFFECT OF MUTATION AND RECOMBINATION ON THE GENOTYPE-PHENOTYPE MAP} 
 
\author{ {\bf C. R. Stephens }\\ 
NNCP, Instituto de Ciencias Nucleares,\\   
UNAM, Circuito Exterior, A.Postal 70-543 \\ 
M\'exico D.F. 04510 \\
e-mail: stephens@nuclecu.unam.mx\\}

\begin{document} 
 
\maketitle 
  
\begin{abstract} 

The effect of genetic operators other than selection, such as mutation 
and recombination, on the genotype-phenotype map is considered. In particular, 
when the genotypic fitness landscape exhibits a ``symmetry'', i.e. many 
genotypes corresponding to the same phenotype have equal fitness values,
it is shown that such operators can break this symmetry. The consequences
of this ``induced symmetry breaking'' are investigated. Specifically, 
it is shown that it generically leads to an increase in order or
self-organization in the system and to the phenomenon of orthogenesis. 
Additionally, it is shown that it potentially leads to a more robust 
evolution circumventing some of the problems of brittleness. The 
above points are supported by explicit, analytic results associated
with some simple one and two-locus models and also by some much more 
complicated numerical simulations.

\end{abstract}

\section{Introduction}

Modulo the debate over the competing roles of selection and mutation
the Darwinian concept of natural selection has stood alone for nearly
a century and a half as the principle source of order in the natural 
world. More recently another paradigm has been presented \cite{kauffman} 
which draws for inspiration on the emergence of order in the physical
rather than the biological world. Simply put: is order a consequence
of the adaptive changes that take place in a system due to the effect of its 
environment, or does order appear ``spontaneously'', irrespective of
any inherent selection? As in the selectionist/neutralist debate 
the  correct answer is that order will appear both 
spontaneously and due to selection. However, for which systems
one predominates over the other is a much more vexed question. 

Traditionally, the tendency has been to view selection as an
ordering agent and mutation and recombination as ``disordering''
effects. The Neutral theory \cite{kimura}, for instance, in its 
traditional guise makes no statement about any adaptive value of genetic 
drift, though others \cite{wright1,maynardsmith,huynen} have raised the issue
of whether or not adaptive evolution can benefit from neutral evolution.
Thus, genetic operators other than selection have generally been discounted
as potential sources of order. Here, I am defining a genetic operator
to be any operation $H$ such that $P(t+1)=HP(t)$, where $P(t)$ is the population
at time $t$. 

In this short paper I will attempt to put other operators, such as
mutation and recombination, onto a more democratic footing vis a vis selection.
by presenting and discussing a third alternative for 
explaining the origin of order in biological systems that
also has its origin in physics --- ``induced symmetry breaking''. 
The ``symmetry'' here referred to is that inherent in the genotype-phenotype
map when it is many-to-one, i.e. many genotypes correspond to the
same phenotypic fitness value. It is of course not new to emphasize
the importance of the genotype-phenotype map in Darwinian evolution, 
see for instance \cite{gatlin,ratner}, however it is new
to show how this map may self-organize and provide a qualitative and
quantitative framework within which this can be understood. 
In particular, we will see how and under what circumstances the phenomenon
of orthogenesis may come about. 

In section 2 I will introduce the concepts of order, symmetry and symmetry
breaking. In section 3 I will give analytic examples of
induced symmetry breaking in the context of some simple one and two-locus models.
In section 4 I will briefly discuss some results found in some much
more non-trivial models and in section 5 I will make some conclusions.

\section{Order, Symmetry and Symmetry Breaking}

I will not go into detail about a precise definition of ``order''. For the 
purposes of this paper its most salient characteristic is the following:
that for a dynamical system with state space $G$ of dimension $D_G$
for late times the system occupies a subspace $U\subset G$ of dimension
$D_U\ll D_G$. Thus, the more ordered a system is the smaller the subspace
into which it dynamically evolves. 

Intuitively, it is clear that selection
will induce order in this sense. For example, in the presence of pure
selection an entire population will eventually order itself around the
optimum present in the initial population. The dynamical attractor in this
case is typically of dimension zero. In the presence of mutation, such as in the 
Eigen model \cite{eigen}, the quasi-species represents the dynamical 
attractor. i.e. if one starts with a disordered random state then the effect
of selection is to arrive at a more ordered state --- the quasi-species.
As is well known, however, for a large class of fitness landscapes there
exists a critical mutation rate above which there is no dynamical reduction
onto a smaller dimension attractor, i.e. selection has its limits.

However, we must first ask what does selection mean? Selection can be most precisely
thought of in terms of fitness and the corresponding notion of a fitness
landscape \cite{wright2}. Fitness, $f_{\ss Q}:\lra R^+$, 
is most naturally defined on the space of phenotypes, $Q$. In conjunction with 
the genotype-phenotype map, $\phi:G\lra Q$, where $G$ is the space of genotypes,
one may define an induced fitness function on the space of genotypes, 
$f_{\ss G}=f_{\ss Q}\circ\phi$. As the genotype-phenotype map is more often 
than not non-injective (many-to-one) the function $f_{\ss G}$ will be degenerate,
many genotypes corresponding to the same fitness value. Thus, fitness
defines an equivalence relation on $G$, many genotypes being equivalent selectively.
A simple example of this would be the standard synonym ``symmetry'' 
of the genetic code. I will therefore refer to the equivalence of a set of genotypes
under the action of selection (i.e. they're all equally fit) as a symmetry. 
Obviously, by definition, selection preserves this symmetry. 
One can see this explicitly, assuming proportional selection as a concrete example,
from the evolution equation for the probability of finding a genotype $C_i$
\be
P(C_i,t+1)={f(C_i)\o \fav}P(C_i,t)\label{evol}
\ee
where $\fav$ is the average population fitness. Considering the same
equation for a genotype $C_j$, where $C_i$ and $C_j$ both correspond to the 
same phenotype and therefore $f(C_i)=f(C_j)$, one sees that
$P(C_i,t)/P(C_j,t)={\rm constant},\ \forall t$. We can in fact take this to be the
defining characteristic of the symmetry: that for ${C_g}\subset G$ where 
$\phi({C_g})=C_q$, $C_q$ being a given phenotype,    
$P(C_i,t)/P(C_j,t)={\rm constant},\ \forall t,\ {\rm and}\ \forall\ C_i, C_j\in{C_g}$.

How may this symmetry be broken? In a finite gene pool the symmetry will
be broken spontaneously by stochastic effects. This can be understood 
in several ways, e.g. via the theory of branching processes \cite{taib} or 
using Kimura's difusion approximation \cite{kimura}. To lend a term from physics, 
such ``spontaneous symmetry breaking''
lies at the heart of Kauffman's ideas about the origin of order.  
Thus, even in the absence of selection a system can dynamically evolve
to a smaller subspace, i.e. spontaneous symmetry breaking can lead to an
increase in order. 

I will now turn to another form of symmetry breaking by considering 
the effect of the other genetic operators besides selection defining 
\be
P(C_i,t+1)=\nn\\ 
H(\{f(C_j)\},\{p_k\},\{P(C_j,t)\},t)P(C_i,t)
\ee
where $H$ is an operator that depends on the fitness landscape, $\{f(C_j)\}$,
the probabilities, $\{p_k\}$, to implement the various genetic operators 
and on the population composition $\{P(C_j,t)\}$. I assume that one can
write $H\equiv H_s+H_o$, where $H_s$ is the part of the evolution operator
associated with pure selection and $H_o$ contains the effect of the 
other genetic operators. The landscape symmetry will thus be preserved by
the action of the other genetic operators if 
$H_oP(C_i,t)=H_oP(C_j,t)\ \forall t,\ {\rm and}\ \forall C_i, C_j\in{C_g}$.
If this condition is not satisfied we will say that the symmetry has
been broken by the action of the other genetic operators; instead
of a spontaneous symmetry breaking there is an ``induced'' symmetry breaking.  

As a quantitative measure of this symmetry breaking we will use
the concept of ``effective'' fitness, defined via \cite{stewael,stewael2}
\be
P(C_i,t+1)={f_{\ss\rm eff}(C_i,t)\o \fav}P(C_i,t)
\ee
One may think of the effective fitness as representing the effect of all
genetic operators in a single ``selection'' factor. Hence, if only 
pure selection was allowed
$f_{\ss\rm eff}(C_i,t)$ would represent the fitness value at time $t$ required 
to increase or decrease $P(C_i,t)$ by the same amount as an evolution
involving all the genetic operators and with selective fitness $f(C_i)$. 
If $f_{\ss\rm eff}(C_i,t)>f(C_i,t)$ then the effect of the genetic
operators other than selection is to enhance the number present of genotype
$C_i$ relative to the number found in the absence of those operators.
The converse is true when $f_{\ss\rm eff}(C_i,t)<f(C_i,t)$.

\section{Analytic Examples of Induced Symmetry Breaking}

We will now illustrate the phenomenon of induced symmetry breaking in
some very simple examples of one and two-locus systems. Consider a
single genetic locus with two alleles, $0$ and $1$ which have the same fitness
value, $f$. In the absence of mutations
$f_{\ss\rm eff}(C_i,t)=f(C_i,t)=f,\ \ \forall i=0,1$
``Synonym'' symmetry here is manifest in the fact that in the infinite
population case $\Delta P(t)=P(1,t)-P(0,t)$ is constant in time. Thus,
any initial deviations from homogeneity in the initial population will
be preserved. For non-zero mutation rate, any initial inhomogeneity will
be eliminated by the effect of mutations. i.e. if $\Delta P(t)>0$ one will
find that $f_{\ss\rm eff}(0,t)>f_{\ss\rm eff}(1,t)\ \forall t$ until the
deviation is eliminated. Hence, one sees that the effect of mutations is to
break the landscape symmetry between alleles $0$ and $1$. This
mutation induced symmetry breaking brings the system into ``equilibrium'',
i.e. into the homogeneous population state. During this approach
to equilibrium the less numerous allele, $0$, is ``selected'' more
than the allele $1$ in that it leaves more offspring. If the mutation rates
for changing allele $1$ to allele $0$ and for changing allele $0$ to allele
$1$ are not equal, but are $p_1$ and $p_2$ respectively, then the
induced symmetry breaking is even more pronounced as can be seen by
\be
\Delta P(t+1)=(1-2p_2)\Delta P(t)+(p_1-p_2)P(0,t)
\ee
In this case $\lim_{t\ra\infty}\Delta P(t)\ra ((p_1-p_2)/(p_1+p_2))$

Now consider a two-loci system, once again with two alleles, $0$ and $1$,
evolving with respect to selection and mutation.
The fitness landscape we will take to be: $f(00)=f(01)=1$, $f(11)=10$, 
$f(10)=0.1$. The fitness landscape in this case is only partially degenerate:
the states $00$ and $01$ having the same fitness value. However, although
the fitness values are the same the effective fitness values are different:
$f_{\ss\rm eff}(00,0)=(1-0.9p+9.9p^2)$, $f_{\ss\rm eff}(01,0)=(1+9p-9.9p^2)$,  
where $p$ is the mutation rate and initial proportions of all four states
are equal at $t=0$. Here, the synonym symmetry is being broken
due to the fact that the fit chromosome $11$ can more easily mutate 
(for $p<0.5$) to the chromosome $01$. Therefore, there is a population flow
away from $00$ to $01$ even though there is zero fitness gradient to cause it.
Thus, we see a tendency for the system to evolve along a preferred direction
not because of selection constraints but because the system has preferred
directions of change in the face of {\it random} mutations. This is the
phenomenon of orthogenesis and is simply a result of induced symmetry 
breaking and is quantitatively measured by the effective fitness function.

Naturally this phenomenon encourages one to ask just when neutral evolution
is actually ``neutral''. In the above case it is not neutral to the presence
of non-neutral adjacent mutants. 
The idea that neutral evolution can facilitate adaptive evolution is not new
\cite{wright1,maynardsmith,huynen}, however a clear, well defined framework 
within which this can be understood, induced symmetry breaking and the concept
of effective fitness, is. In fact, it is clear that neutral evolution 
precisely leads to adaptive evolution when the effective fitness landscape
is non-flat. For a flat fitness landscape where all strings have fitness $f$
\be
f_{\ss\rm eff}(C_i,t)=f\sum_{j=1}^{2^N-1}{P(C_j,t)\o P(C_i,t)}
p^{d_{ij}}(1-p)^{N-d_{ij}}
\ee
where $d_{ij}$ is the Hamming distance between the strings $C_i$ and $C_j$.
For a homogeneous population the number of states Hamming distance $d_{ij}$
from $C_i$ is ${}^NC_{d_{ij}}$ thus $f_{\ss\rm eff}(C_j,t)=f\ \forall C_j,t$.
Thus, under these circumstances the effective fitness landscape is as flat
as the normal one and there is no symmetry breaking. Small deviations from 
homogeneity will be manifest in small corrugations of the effective fitness
landscape which will gradually diminish as the population homogenizes. If the
landscape only has a flat subspace then how well one can describe the 
population evolution as being neutral will depend on where the population
is located and, if located predominantly in the flat subspace, what is the
Hamming distance to states not within the subspace and what is the fitness
of those states. Pictorially, if one thinks of a bowl with a flat bottom
then the sides of the bowl with the largest gradient will attract the 
population most strongly. 
 
Above I considered only mutation as a source of induced symmetry breaking.
Similar considerations apply also to recombination. 
For the two-locus system mentioned above 
$f_{\ss\rm eff}(00,0)=(1-(9.9p_c/12.1))$ and $f_{\ss\rm eff}(01,0)=(1+(9.9p_c/12.1))$  
where $p_c$ is the recombination probability. Thus, once again we see
the landscape symmetry broken by the effects of another genetic operator.
A simple, but striking example of induced symmetry breaking can be seen with the 
landscape of the well known counting ones, or unitation problem. A 
population of $5000$ $8$-bit strings was considered. 
Figure 1 is a plot of $M(l)$ versus time where
$M(l)\equiv (n_{\ss opt}(l)-n_{\ss opt}(8))/n_{\ss opt}(8)$. Here, 
$n_{\ss opt}(l)$ is the number of optimal $2$-schemata of defining length $l$ 
normalized by the total number of length $l$ 2-schemata per string, i.e. $9-l$. 
By optimal $2$-schemata we mean schemata containing the global optimum $11$.
$n_{\ss opt}(8)$ is the number of optimal $2$-schemata of defining length $8$.
Figure 1 is with $p_c=1$. Averages over $30$ different runs are shown. 
In terms of fitness there is absolutely no preference for one size of
optimal two-schema versus another, however, recombination breaks this symmetry
in a very dramatic fashion giving a preference for long rather than short schemata.
\begin{figure}[h]
$$
\psfig{figure=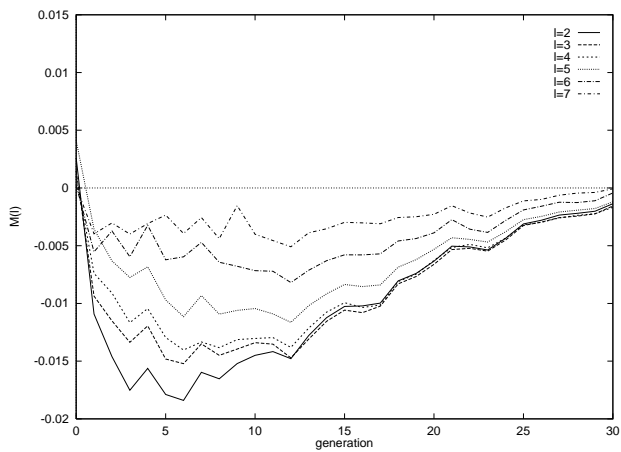,angle=-90,width=3.3in}
$$
\caption{Graph of $M(l)$ versus $t$ in unitation model with $p_c=1$.}
\end{figure}

\section{Numerical Examples}

In the previous section I used some very simple tractable models to
illustrate the phenomenon of induced symmetry breaking. In this section
I will present some more non-trivial examples. For more details I refer
the reader to the original articles. 

\par\noindent i) Self-Adaptation: It is well known that mutation and recombination
rates are not uniform throughout a structure such as a protein. One may
well wonder why certain values are found rather than others and if or not
there is any adaptive value in it. In fact, in the case of the HIV virus it
can be shown that preference for non-synonymous mutations in the neutralization
epitope of the virus is directly due to an induced symmetry breaking \cite{codon}.

Normally one thinks of the mutation and recombination rates as exogeneous 
parameters. However, if one considers a system where they are coded in the
chromosome, but are not directly selected for, then one has a completely 
autonomous system wherein one may examine whether the mutation and 
recombination rates across the population exhibit any degree of self-organization.
More explicitly, coding the two rates into an $N_c$-bit extension of a chromosome 
of length $N$ which represents a non-degenerate fitness landscape leads to a 
new one which has a degree of degeneracy of $2^{N_c}$, i.e. the phenotype-genotype
map is $2^{N_c}$ fold degenerate. In practice, starting off with a random population
where the average rates are $0.5$ one finds that the population in a class of 
interesting landscapes self-organizes until preferred mutation and recombination
rates appear \cite{artlif}.
It is important to emphasize that such self-organization cannot come about as 
a consequence of selection, as by construction mutation and recombination rates
are not selected for. However, the genetic operators of mutation and recombination
themselves break the symmetry. The effective fitness measures the strength of this
induced symmetry breaking is.  

As a specific example, consider a time dependent landscape defined on 
6-bit chromosomes 
that code the integers between $0$ and $63$, where the initial landscape
has a global optimum situated at $10$ and $11$ and a local optimum at $40$ and $41$. 
However, after 60\% of the 
population reaches the global optimum the landscape is suddenly changed to a new landscape 
wherein the original global optimum is now only a local optimum.  The original local
optimum at $40$ and $41$ remains the same but with a higher fitness value than the 
new local optimum at $10$ and $11$ and furthermore a new
global optimum appears at $63$. I will denote this landscape the ``jumper'' landscape. 
Figure 2 shows the results of an experiment where the mutation and crossover probabilities
were coded in the chromosomes, either with three or eight bits to codify each
probability. Tournament selection of size 5 was used and a lower bound of $0.005$ 
for mutation imposed. The success of the self-adapting system in converging to the
time dependent global optimum was compared to that of an ``optimal'' fixed 
parameter system with $p=0.01$ and $p_c=0.8$.
\begin{figure}[h]
$$
\psfig{figure=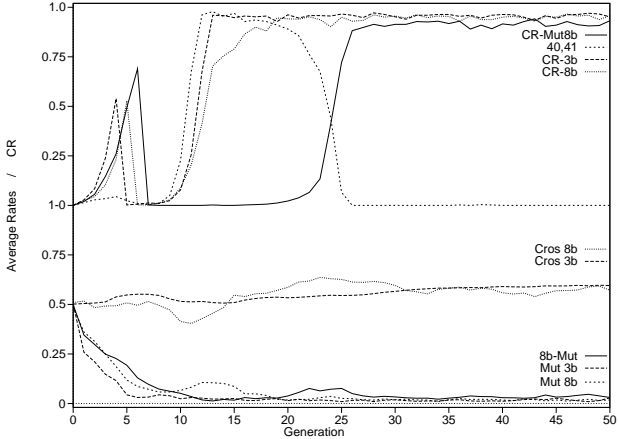,angle=-90,width=3.3in}
$$
\caption{Graph of relative concentration of the global optimum (CR) (upper graph)
and average crossover and mutation probabilities (lower graph) 
as a function of time for the ``jumper'' landscape. 
CR-3b and CR-8b are the results for $3$-bit and $8$-bit
encoded algorithms. CR-Mut8b is the result for coded mutation with $p_c=0$, 
with $40,41$ being the relative concentration of strings associated with the local
optimum at $40$ and $41$. Mut 3b, Mut 8b, Cross 3b and Cross 8b are the average
mutation and crossover probabilities in $3$-bit and $8$-bit representations. The
solid line for Mut8b is the average mutation rate in the case $p_c=0$.}
\end{figure}

The upper curves show the relative frequencies of the optima using $8$-bit and 
$3$-bit codification and also 
what happens when $p_c=0$ and only the mutation rate is coded. There are several 
notable features: first of all, the optimal fixed parameter system was 
incapable of finding the new optimum whereas the coded system had no such problem.  
For the case $p_c=0$ the curve $40,41$ shows the relative frequency
of the strings associated with the optimum at $40$ and $41$. Before the landscape
``jump'' this optimum is local being less fit than the global optimum 
at $10$ and $11$. After the ``jump'' it is fitter but less fit than the new 
global optimum $63$ which is an isolated point. 

One thus sees that
the global optimum was found in a two-step process after the landscape change. 
First the
strings started finding the optima $40$, $41$ before moving onto the true global
optimum, $63$. Immediately after the jump the effective population of the new
global optimum is essentially zero. The number of strings associated with 
$40$ and $41$ first starts to grow substantially at the expense of $10$ and
$11$ strings. At its maximum the number of optimum strings is still very low,
however, very soon thereafter the algorithm manages to find the optimum string which
then increases very rapidly at the expense of the rest. The striking result
here can be seen by comparing the changes in the relative frequencies with
the changes in the average mutation rate, especially in the case $p_c=0$. 
Clearly they are highly correlated.
First, while the population is ordering itself around the original optimum,
there is an effective selection against high mutation rates as one can see
by the steady decay of the average mutation rate. After the jump there
is a noticeable increase in the mutation probability as the system now has to 
try to find fitter strings. As the global optimum is an isolated state it is much 
easier to find fit strings associated with $41$ and $40$. The population is
now concentrated on this local optimum and starts to cool down again only to find that this 
is not the global optimum, whereupon the system heats up again to aid the removal of the
population to the true global optimum. It is clear that there is a small delay
between the population changes and changes in the mutation rate. This 
is only to be expected given that there is no direct selective advantage 
in a given generation for a particular mutation rate. The selective advantage
of a more mutable genotype over a less mutable one can only come about via a feedback mechanism.
It is precisly this feedback process that is described and measured by the 
effective fitness function. 

The average mutation rate also grows due to another
effect which is that the new optimum is more likely to be reached by strings
with high mutation rates which then grow strongly due to their selective advantage. 
Thus, high $p$ strings will naturally dominate the early evolution of the global
optimum. After finding the optimum however it will become disadvantageous
to have a high mutation rate hence low mutation strings will begin to 
dominate. Induced symmetry breaking here is once again manifest in a most
striking way. Although there is no direct selective benefit to differently
coded strings their ability to produce offspring that can adapt to the 
changing landscape is very different.

\par\noindent ii) Neuro-genetic models: In this case an analysis was made 
\cite{angeles} of the population dynamics of a variant of Kitano's 
neurogenetic model \cite{kitano1,kitano2} wherein the chromosome encodes the rules 
for cellular division and the phenotype is a 16-cell organism interpreted 
as a connectivity matrix for a feedforward neural network. Specifically, 
an artificial ecological environment was studied which 
consists of a single species composed of neural networks as individuals. 
Every chromosome, or genotype, is used to produce a particular architecture 
for a feedforward NN that consists of $12$ input neurons, $4$ hidden and $1$
output neuron --- the phenotype. A genetic algorithm is then applied  
to the chromosomes present in the population at each epoch which induces a  
search of the connectivity matrix space determined by the structure of the NN. 
Environmental effects are included in the fitness function that 
measures the learning capacity of a particular individual. 

A chromosome consists of eight blocks of four genes each one of which is 
a three bit structure. The blocks themselves are labelled from 
{\it a} to {\it h}. The reproduction process 
always begins with block {\it a}. Thus the first four genes have a priviliged role 
as they label the cells that are going to be reproduced 
in the second step of reproduction. As an example consider the chromosome
{\it baea.dcaa.defa.becd.aaea.aafh.haec.fgaa}. The two step reproduction process
specified by this chromosome can be written 
\be
a \lra \pmatrix{
b & a\cr
e & a\cr}\lra
\pmatrix{
d & c & b & a\cr
a & a & e & a\cr
a & a & b & a\cr
e & a & e & a\cr
}\nn\\
\leftrightarrow \pmatrix{
0 & 1 & 1 & 0 & 1 & 0 & 0 & 0 & 1 & 0 & 0 & 0\cr
0 & 0 & 0 & 0 & 0 & 0 & 1 & 0 & 0 & 0 & 0 & 0\cr
0 & 0 & 0 & 0 & 0 & 0 & 0 & 0 & 1 & 0 & 0 & 0\cr
1 & 0 & 0 & 0 & 0 & 0 & 1 & 0 & 0 & 0 & 0 & 0\cr}
\ee
Thus the first block, {\it baea}, codes for the division of the original cell 
{\it a} into four cells. The first of these cells, {\it b},  then divides into 
four more which form the 
upper left quadrant, {\it dcaa}, of the matrix. The second cell, {\it a}, 
maps block {\it a} of the chromosome into the upper right quadrant etc. 
Finally, one constructs the connectivity matrix by reading left to right, row by row. 
Thus a $1$ specifies a connection between an input neuron
and a hidden neuron and a $0$ its absence.

The genotype-phenotype map in this case is highly degnerate.
For example, in the above we can change blocks {\it c}, {\it e}, {\it f}, {\it g} and {\it h}
without changing the resulting phenotype. It is also a non-local 
function on the chromosomes since entries of block number one can
target any one of the other blocks irrespective of their distance.
To define a fitness function the 
learning speed of the NNs on a given test function was measured
$$y_c={\epsilon \over 3} (x_1+x_2+x_3)  + (1-\epsilon) X$$
where $\epsilon$ is a noise control parameter and $X$ is a randomly
generated number. A genetic algorithm was used to search the space of network
architectures for the one capable of learning this function with the smallest 
number of attempts. Given the highly degenerate nature of the 
genotype-phenotype map one might expect to see an optimum phenotype emerge
corresponding to a random distribution of corresponding genotypes. However,
this was not the case --- certain genotypes were consistently preferred
thus indicating that the genotype-phenotype symmetry was broken. The reason for this 
is that although degenerate genotypes were equivalent selectively the
other genetic operators, mutation and recombination, broke the 
symmetry picking out those genotypes best able to withstand the 
effects of mutation and recombination, i.e. those that were most
likely to lead to other ``fit'' neural networks. Remarkably,
it was found that the induced symmetry breaking in this context
could be described in terms of the emergence of an ``algorithmic
language'' \cite{angeles}. 

\par\noindent iii) Giraffe necks \cite{virus}: 
This model consists of a population of one thousand genotypes 
subject to random mutations. A genotype is a cellular automata with
binary elements which gives rise to a giraffe neck size, i.e. a phenotype, 
given by the 
number of automata elements that are ``switched on" at the  
fixed point (steady state) of the automata dynamics. As there are 
many different automata that can evolve to the same fixed point the 
genotype-phenotype mapping is highly degenerate. One ``master'' 
gene in particular plays a special role as it governs the way in which the
Boolean rules used in the evolution mutate. 

Each member of the population is selected for the next generation with
probability $P_i = f_i / \sum_j f_j$, where $f_i$ is the fitness
of phenotype $i$. Initially, there are ample resources available from both 
small and large trees, the only selective criterion being that
giraffes prefer to choose a mate among those that have similar
neck size. This ``social pressure" landscape is modelled by defining the
fitness of the $ i{\rm th} $ giraffe to be a function of its neck size 
$ n_i $ and the average neck size of the population, $ \left< n \right> $,
with value one if $ \left< n \right> -
\delta < n_i < \left< n \right> + \delta $ and zero otherwise.
Here, $ \delta > 0 $ is a tolerance window. Note that landscape
fitness depends only on neck size, hence all genotypes that correspond
to the same dynamical fixed point (phenotype) have the same
fitness. Thus there is no direct selective advantage for one genotype
versus another.  To introduce time dependence into the landscape one
imposes a short period of drought in which food begins to be
available only in taller and taller trees. This period is mimicked by
making $ f_i = 1 $ if $\left< n \right> - \delta + \epsilon <
n_i < \left< n \right> + \delta + \epsilon $, where $ \epsilon $ is a
stress parameter, and zero otherwise. After this drought the landscape
is restored to its original state.

The ``master'' gene divides the population into two genetic categories, type
zero and type one, which can mutate one into the
other due to the effect of purely random mutations that have a probability 
$\mu$, except for the master gene which mutates at a rate $\nu$. 
Type zero chromosomes, by nature of the dynamical evolution
rules they are associated with, tend to
give rise to giraffe offspring with shorter necks, whilst type one chromosomes,
when they are expressed, tend to lead to giraffes with longer necks. Before
the draught there is a period in which type one is not expressed. After
a certain period of time it becomes expressed then afterwards the draught starts.
The social pressure landscape implies there are two possible attractors: all
type one or all type zero. The effect of the drought is to change between one
and the other.

A typical experiment leads to the following results, the general behaviour can
be seen in Figure 3:
\begin{figure}[h]
$$
\psfig{figure=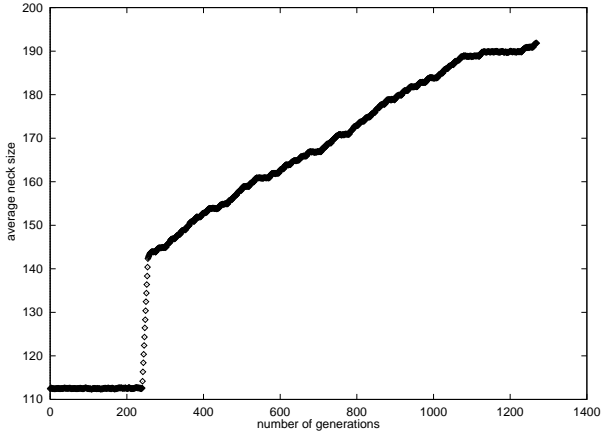,angle=-90,width=3.3in}
$$
\caption{ Graph of average giraffe neck size (in arbitrary units) as
a function of time for a population of 1000 giraffes. The drought starts at
generation 240 and lasts for 30 generations. Subsequent neck growth lasts
for another 1000 generations. The 
parameter values used were: $ \mu = 0.0025 $, $ \nu = 10^{-6} $, $ \delta =
2.0 $ and $ \epsilon = 1.0 $.}
\end{figure}
In the initial period of evolution, before the drought, average neck size is short.
After the drought arrives the average neck size grows very quickly.
After it ends it continues to grow, albeit more slowly, for
a substantial amount of time until a steady state is reached.  
These results can be explained quite simply:
In the period before the draught, and before expression, type one chromosomes
increase due to the effect of neutral drift. After expression they are
effectively selected against due to their tendency to produce giraffes with
longer necks that pass outside the tolerance threshold and therefore cannot 
reproduce. Thus, before the drought the effective fitness of type one chromosomes
is low. However, due to the effect of mutations type one chromosomes 
are not eliminated totally but constitute about $1- 5\%$ of the total population.
After the draught starts the effective fitness of type one chromosomes
increases substantially, given that they lead to giraffes of longer necks. The
result is that the population becomes dominated by type one chromosomes, 
with a small fraction of type zero remaining due to the effects of mutation.
After the end of the drought as type one chromosomes tend to produce longer
necks the average neck size increases until a steady state is reached and
it cannot grow anymore. 

In the giraffe model there
is absolutely no direct selective difference between type one and type zero
chromosomes. The only advantage of one versus the other is in how they
produce well adapted offspring, a quantitative measure of this being the 
effective fitness.

\section{Conclusions}

In this contribution I have tried to briefly lay out the case for induced 
symmetry breaking as an origin of order in biological systems. Without 
doubt it exists, as has been conclusively demonstrated. It is possible to
see it at work in simple analytic one and two locus population models,
and also numerically in several much more non-trivial examples of artificial
life system as I have briefly touched upon here. The extent to which 
it exists in real biological systems is a question for future research. 
The chief difficulty in applying these ideas to the latter is that it
is very difficult to assure oneself that apparent selection for a particular 
genotype is due to an effective selection, via a symmetry breaking effect,
and not via some direct, yet unobserved, selective factor. For this reason I 
believe it is well worthwhile continuing with the examination of 
mathematical models of increasing complexity wherein one may better
control the fitness landscape and the nature of the genotype-phenotype map,
and also to consider artificial life systems where there is much more
control over selective factors.

One might enquire as to why bother introducing the concepts of effective fitness
and induced symmetry breaking. There are several reasons: first of all
they allow one in a quantitative sense to understand the different mechanisms
by which order may arise in biological systems. Secondly, they provide a 
framework within which neutral evolution and natural selection can be
understood as different sides of the same coin, and in particular under
what circumstances neutral mutations may lead to adaptive changes. 
Thirdly, induced symmetry breaking may well lead to more robust 
adaptive systems. It is no good having an extremely fit phenotype if when
subjected to mutation at the genotypic level it typically mutates into
an unfit phenotype. Rather one requires that an organism not only be fit
but that it gives rise to fit offspring which in their turn give rise to fit
offspring etc. Induced symmetry breaking can pick out precisely those
evolutionary pathways that possess this type of robustness as is
found in the neurogenetic model of section 4. 

To what extent the different possible sources of order predominate will
depend very much on the landscape considered and is as open to debate
as the standard selectionist/neutralist argument. I believe that artificial
life research can play an important role in this debate by examining
the generic properties of landscapes and populations that admit as dominant one source 
of order versus another.

\subsubsection*{Acknowledgements} 

This work was partially supported through DGAPA grant number IN105197.
The basic idea of induced symmetry breaking was developed in collaboration 
with Henri Waelbroeck to whom the author is grateful for many stimulating
conversations.
 
\subsubsection*{}

\end{document}